%% file: domwall.tex
\documentclass[prl,superscriptaddress,twocolumn,floatfix]{revtex4-1}
\usepackage{graphicx,color}
\usepackage{amsmath,amssymb,bm}
\usepackage{braket}

\newcommand{\xidw}{\xi_\mathrm{dw}}
\newcommand{\VARn}{\mathrm{VAR}_n}
\newcommand{\chat}{\hat{c}^{\phantom{\dagger}}}
\newcommand{\chatdag}{\hat{c}^{\dagger}}
\newcommand{\nhat}{\hat{n}}

\begin{document}
\title{Domain-wall melting as a probe of many-body localization}

\author{Johannes Hauschild}
\email[E-mail: ]{johannes@pks.mpg.de}
\affiliation{Max Planck Institute for the Physics of Complex Systems, D-01187 Dresden, Germany}
\author{Fabian Heidrich-Meisner}
\affiliation{Department of Physics and Arnold Sommerfeld Center for Theoretical Physics, Ludwig-Maximilians-Universit\"at M\"unchen, D-80333 M\"unchen, Germany}
\author{Frank Pollmann}
\affiliation{Max Planck Institute for the Physics of Complex Systems, D-01187 Dresden, Germany}

\begin{abstract}
	Motivated by a recent optical-lattice experiment by Choi {\it et al.}~[Science {\bf 352}, 1547 (2016)], we discuss how domain-wall melting can be used to investigate many-body localization. First, by considering noninteracting fermion models, we demonstrate that experimentally accessible measures are sensitive to localization and can thus be used to detect  the delocalization-localization transition, including divergences of characteristic length scales.
Second, using extensive time-dependent density matrix renormalization group simulations, we study fermions with repulsive interactions on a chain and a two-leg ladder. The extracted critical disorder strengths agree well with the ones found in existing literature. 
\end{abstract}


\maketitle
{\bf Introduction.} In pioneering works based on perturbation theory \cite{Basko06,Gornyi2005}, it was shown  that Anderson localization,
i.e., perfectly insulating behavior even at finite temperatures,
can persist in the presence of interactions.
Subsequent theoretical studies on mostly one-dimensional (1D) model systems have
unveiled many fascinating properties of such a \emph{many-body localized} (MBL) phase. The MBL phase is a dynamical
phase of matter defined in terms of the properties of highly excited many-body eigenstates. It is characterized by
an area-law entanglement scaling in all eigenstates \cite{Bauer2013,Kjaell2014,Luitz2015}, a logarithmic increase of
entanglement in global quantum quenches \cite{Bardarson2012,Znidaric2008,Serbyn2013}, failure of the eigenstate
thermalization hypothesis \cite{Pal2010} and therefore, memory of initial conditions \cite{Altman2015,Nandkishore2015}.
The phenomenology of MBL systems is connected to the existence of
a complete set of commuting (quasi) local integrals of motion (so-called ``l-bits'') that are believed to exist in systems  in which all many-body eigenstates are localized \cite{Huse:2014uy,Serbyn2013,Chandran:2015df}. These l-bits can be thought of as quasiparticles with an infinite lifetime, in close analogy to a zero-temperature Fermi liquid \cite{Bera2015,Basko06}.
Important open questions pertain to the nature of the MBL transition and the existence of an MBL phase in higher
dimensions, for which there are only few results (see, e.g., \cite{Inglis2016,BarLev2016}),
mainly due to the fact that numerical simulations are extremely challenging in dimensions higher than one for the MBL problem.

The phenomenology of the MBL phase has mostly been established for closed quantum systems. A sufficiently strong coupling of a disordered, interacting system to a bath is expected to lead to thermalization (see, e.g., \cite{Nandkishore2014,Johri2015}). Thus, the most promising candidate systems for the experimental investigation of MBL physics are quantum simulators such as ultracold quantum gases in optical lattices or ion traps. 
So far, the cleanest evidence for MBL in an experiment has been reported for an interacting  Fermi gas in an optical  lattice with quasiperiodicity, realizing the Aubry-Andr{\'e} model \cite{Schreiber2015,Bordia2015}. 
Other quantum gas experiments used the same quasi-periodic lattices or laser speckles to investigate Anderson localization \cite{Billy2008,Roati2008} and the effect of interactions \cite{DErrico2014}, however, at low energy densities. Experiments with ion traps provide an alternative route, yet there, at most a dozen of ions can currently be studied \cite{Smith2015}.

\begin{figure}[b]
\includegraphics[width=\columnwidth]{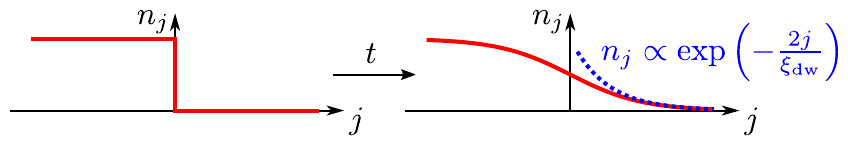}
\caption{Initial state (left) and density profile after a sufficiently long time (right) in the
        localized regime. The profile develops an exponential decay with distance
        $n_j \propto \exp(- \frac{2 j}{\xidw})$ in its tails away from the initial edge $j=0$ of the domain wall.
}
\label{fig:sketch}
\end{figure}

By using a novel experimental approach, a first demonstration and characterization of MBL in a two-dimensional (2D)
optical-lattice system of interacting bosons with disorder has been presented by Choi {\it et al.}~\cite{Choi16}. They
start from a state
that contains particles in only one half of the system while the rest is empty. Once tunneling is allowed, the particles from the initially
occupied region can spread out into the empty region (see Fig.~\ref{fig:sketch}). The evolution of the particle density is  tracked using
single-site resolution techniques \cite{Bakr2009,Sherson2010} and digital mirror devices allow one to tune the disorder. 
The relaxation dynamics provides  evidence for  
the existence of an ergodic and an MBL regime as disorder strength is varied, characterized via several 
observables such as density profiles, particle-number imbalances and measures of the localization length \cite{Choi16}.
This experiment serves as the main motivation for our theoretical work.

The term domain-wall melting is inherited from the equivalent problem in quantum magnetism (see, e.g.,
\cite{Antal1999,Gobert2005,Steinigeweg2006,Lancaster2010,eisler2013,Santos2008}), corresponding to coupling two ferromagnetic domains with opposite spin orientation. 
Furthermore, the domain-wall melting describes the transient dynamics \cite{Vidmar2013,Hauschild2015,vidmar15} of  sudden-expansion experiments of interacting quantum gases in optical lattices 
(i.e., the release of initially trapped particles into an empty homogeneous lattice) \cite{schneider12,ronzheimer13,Xia2015,vidmar15}.
Theoretically, the sudden expansion of interacting bosons in the presence of disorder was studied in, e.g., \cite{Roux2008,Ribeiro2013}  for the expansion
from the {\it correlated ground state} in the trap, while for MBL, higher energy densities are  relevant.

We use exact diagonalization (ED) and time-dependent density matrix renormalization group (tDMRG) methods \cite{White2004,Vidal2004,Daley2004,Zaletel2014} to clarify some key questions of the domain-wall experiments. First, by considering noninteracting fermions in a 1D tight-binding model with diagonal disorder we demonstrate that it is possible to extract the single-particle localization length $\xi^{(1)}_{\rm loc}$ as a function of disorder strength from such an experiment
since the density profiles develop exponential tails with a length scale $\xidw$  (see
Fig.~\ref{fig:sketch}). This domain-wall decay length 
$\xidw$ also captures the  disorder driven metal-insulator transition in the Aubry-Andr{\'e} model when approached from the localized regime, exhibiting a divergence. Second, we study the case of  spinless fermions 
with nearest-neighbor repulsive interactions on chains and two-leg ladders, for which numerical estimates
of the critical disorder strength $W_c$ of the metal-insulator transition
are available \cite{Pal2010,Oganesyan2007,Luitz2015,BarLev2015,Bera2015,Devakul2015,Baygan15}. 
For both models, essential features of the noninteracting case carry over, namely, the 
steady-state profiles decay exponentially with distance in the localized regime $W>W_c$ (i.e., the expansion stops), while particles continue
to spread in the ergodic regime $W< W_c$. Moreover, we discuss experimentally accessible measures to investigate the dynamics close to the 
transition for all models. 

{\bf Noninteracting cases.} We start by considering fermions in a 1D tight-binding lattice with uncorrelated diagonal disorder. The Hamiltonian
reads:
\begin{equation}
	H_0 = - \frac{J}{2}\sum_{\langle i, j\rangle} ( \chatdag_i \chat_j + H.c.)
		- \sum_j \epsilon_j \nhat_j\,,
	 \label{eq:ham}
\end{equation}
where $\chatdag_j$ denotes the creation operator on site $j$, $\nhat_j = \chatdag_j \chat_j$ is the number operator, $n_j = \langle \nhat_j \rangle$ is density,  and 
$ \epsilon_j \in [-W, W] $ is a random onsite potential ($L$ is the number of sites).  We set the lattice spacing to unity and $\hbar =1$. 
All single-particle eigenstates are localized for any nonzero $W$ and thus the system is an Anderson
insulator at all energy densities \cite{Kramer93, Evers2008}.

\begin{figure}
\includegraphics[width=\columnwidth]{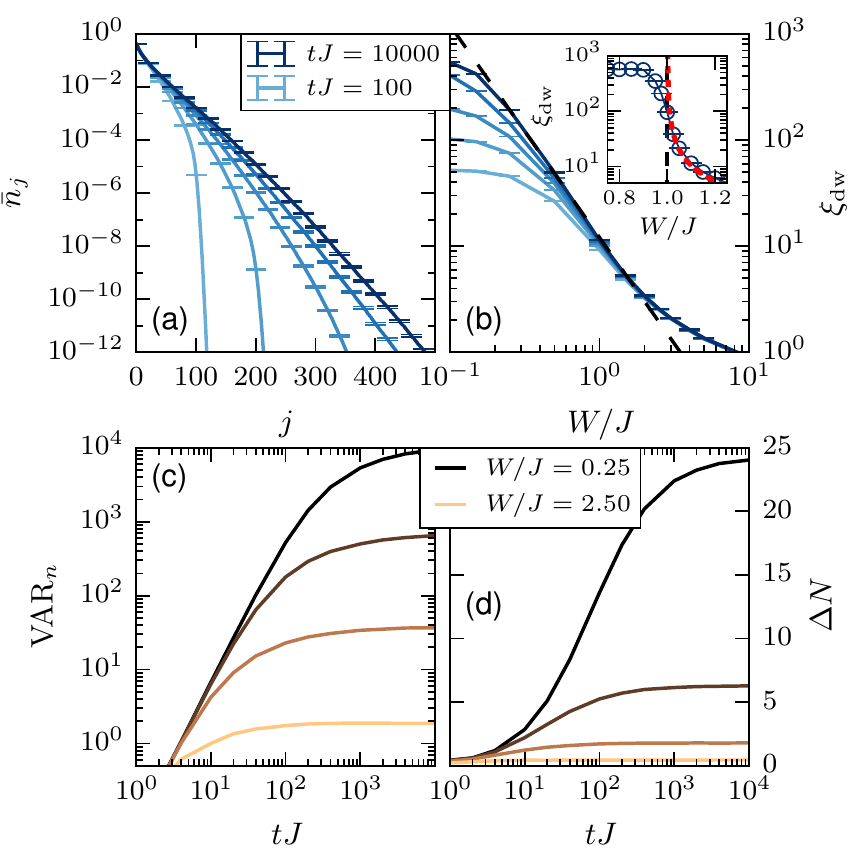}
\caption{ED results ($L = 2000$) for 1D noninteracting fermions  with
	uncorrelated diagonal disorder Eq.~\eqref{eq:ham}. 
	(a) Representative {\em typical} density profile \cite{suppmat} for  $W= 0.5 J$ for times $tJ= 100, 200, 400, 1000, 10\,000$ (bottom to top). 
	(b) Domain-wall decay length $\xidw$ (extracted from $\VARn$) for the same times as in (a), as a function of the disorder strength $W$.
	The dashed line shows a fit to the expected scaling $\xidw \propto W^{-2}$ \cite{Kramer93}.
	(c) Variance  $\VARn$ of the distribution of expanded particles for $W/J=0.25, 0.5, 1.0, 2.5$ (top to bottom).
	Error bars are smaller than symbol sizes and omitted.
	(d) Number of emitted particles $\Delta N(t)$.
	Inset in (b): $\xidw$ at $tJ=10\,000$ (circles) for the Aubry-Andr{\'e} model \cite{suppmat}, which has a delocalization-localization transition at $W=J$,
	compared with the analytical result $\xi^{(1)}_{\rm loc}  = 1/\log\left(\frac{W}{J}\right)$ (red dotted line) \cite{Aubry80}.
}
\label{fig:nonint}
\end{figure}

Typical density profiles for the dynamics starting from a domain-wall initial state are shown for different times in Fig.~\ref{fig:nonint}(a).
Here, ``typical'' refers to the geometric mean $\bar{n}_j$ over disorder realizations (i.e., the arithmetic
mean of $\log n_j$) (see \cite{suppmat}). 
The domain wall first melts slightly yet ultimately stops expanding. 
The profiles clearly develop an exponential tail $\bar{n}_j \propto \exp(- 2 j /\xidw)$ for $j \gg 0$.
The crucial question is now whether the length scale $\xidw$ is directly related to the single-particle localization length or not.

We compare two ways of extracting $\xidw$: First, a fit to the numerical data for $\bar{n}_j$ in the tails $j\gg 0$ and second,
via computing the variance of the particles emitted into the originally empty region.  
For the latter, we view the density $n_j$ in the initially empty region $j>0$ as a spatial distribution
$\langle \cdot \rangle_n \equiv \left( \sum_{j>0} n_j \, \cdot \right) / \Delta N $ where $\Delta N = \sum_{j>0} n_j$ is the number of emitted particles.
The variance $\VARn = \langle j^2 \rangle_n - \langle j\rangle^2_n$ of this particle distribution is shown in Fig.~\ref{fig:nonint}(c) 
and approaches a stationary regime on a timescale depending on $W$. For the time window plotted, only the curves with
$W\geq J$ saturate, yet we checked that also the curves for $W<J$ saturate at sufficiently long times.
At short times, $\VARn \propto t^2$ signals a ballistic expansion of the particles as long as $\VARn(t) \ll \xi^{(1)}_{\rm loc}$.

Assuming a strictly exponential distribution
$n_j \propto \exp(-\frac{2 j}{\xidw})$ for all $j>0$  yields $\VARn \approx \frac{\xidw^2}{4}$ for $\VARn \gg 1$. We use that relation
to extract  
$\xidw$ in the general case as well and in addition, we introduce an explicit time dependence of $\xidw$ to illustrate the approach to the 
stationary state. In general, this gives only a lower bound to $W_c$ since $\VARn$ can be finite for diverging $\xidw$
if the distribution is not exponential. Yet we find that both methods give similar results for the final profile
and show only $\xidw$ extracted from $\VARn$ in Fig.~\ref{fig:nonint}(b).

The known result  for the localization length in the 1D Anderson model 
is $\xi^{(1)}_{\rm loc} = \frac{8(J^2-E^2)}{W^2}$ \cite{Kramer93} for $E=0$ (our initial state leads to that 
average energy for sufficiently large systems). 
Our data for $\xidw$ shown in Fig.~\ref{fig:nonint}(b) clearly exhibit the expected scaling $\xidw \propto W^{-2}$ over a wide range of $W$
as suggested by a fit of $\xidw = a/W^{-2}$ to the data [dashed line in Fig.~\ref{fig:nonint}(b); the prefactor is larger by about a factor of 1.5 than 
the typical localization length  $\xi^{(1)}_{\rm loc}$].
Deviations from the $W^{-2}$ dependence  at small $W$, where $\xi^{(1)}_{\rm loc} \sim  \mathcal{O}(L)$, are due to the finite system size.
At large $W$, the discreteness of the lattice makes it impossible to resolve $\xidw$ that are much smaller than the lattice spacing.
We stress that fairly long times need to be reached to observe a good quantitative agreement with the $W^{-2}$ dependence.
For instance, for  the parameters of Fig.~\ref{fig:nonint}(a), $t J\sim
1000$ is necessary  to reach the asymptotic form.  Nevertheless, even at shorter times, the density profiles are already
approximately exponential.
To summarize, our results  demonstrate that the characteristic length scale
$\xidw$ is a measure of  the single-particle localization length, most importantly 
exhibiting the  same qualitative behavior.  

In Fig.~\ref{fig:nonint}(d), we introduce an alternative indicator of localization, namely, the number of emitted particles $\Delta N(t)$ that
have propagated across the edge $j=0$ of the initial domain wall at a time $t$.
Due to particle conservation, $\Delta N$ is directly related to the
imbalance $\mathcal{I} = \frac{N-2 \Delta N}{N} $ analyzed in the experiment \cite{Choi16}. 
We observe that $\Delta N$ 
shares  qualitatively the same behavior with $\VARn$ [note the linear $y$ scale in Fig.~\ref{fig:nonint}(d)], which will  also apply to  the models discussed in the following.

As a further test, we study the Aubry-Andr{\'e} model in the Appendix \cite{suppmat}.
The comparison of $\xidw$ with the exactly known single-particle localization length \cite{Aubry80} in the inset of Fig.~\ref{fig:nonint}(b)
demonstrates that the domain-wall melting can resolve the delocalization-localization transition at $W = J$.

{\bf Interacting fermions on a chain.}
Given the encouraging results discussed above, we move on to studying the dynamics in a system with an MBL phase,  
namely to the model of spinless fermions with repulsive nearest-neighbor interactions $H_{\rm int} = H_0 + V
\sum_{\langle i,j \rangle} \hat n_i \hat n_{j}$,
 equivalent to the spin-1/2 $XXZ$ chain.
We focus on $SU(2)$ symmetric exchange, i.e., $V=J$, for which numerical studies predict a delocalization-localization transition from an ergodic to the MBL phase at $W_c/ J = (3.5\pm 1)  $ \cite{Luitz2015,Pal2010,BarLev2015,Bera2015} at energy densities in the middle of the many-body spectrum (corresponding to infinite temperature when approaching the transition from the ergodic side). Note, though, that even for this much studied model, some aspects of the phase diagram are still debated in the recent literature
(see, e.g., \cite{Chandran2016,DeRoeck2016}).

\begin{figure}
\includegraphics[width=\columnwidth]{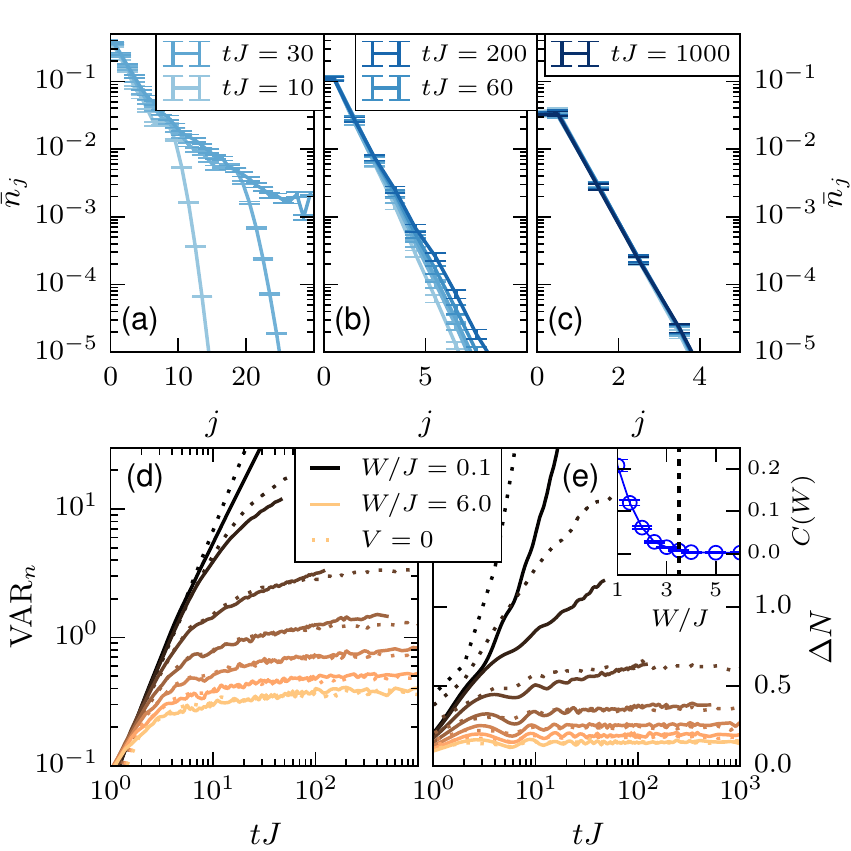}
\caption{tDMRG results ($L=60$) for a chain of interacting spinless fermions with $V=J$. 
	Top row: Typical density profiles for (a) $ W/J=0.5$ at $tJ=10, 20, 30$ (bottom to top), (b) $W/J=3$ and additional data for $tJ= 60, 200$, and
	(c) $W/J=6$, additional data for $tJ=1000$ (on top of the data for shorter times).
	(d) Variance $\VARn$ of the distribution of expanded particles for $W/J=0.1,1,2,3,4,5,6$ (solid lines top to bottom).
	The dotted lines show equivalent data for the noninteracting case $V=0$.
	Error bars are smaller than symbol sizes and omitted.
	(e) Number of emitted particles $\Delta N(t)$.
	Inset: $C(W)$ from fit of $\Delta N(t) $ to Eq.~\eqref{eq:fitDN} for $t J > 10$.
}
\label{fig:HCB_chain}
\end{figure}
Typical time evolutions of density profiles in the ergodic and
MBL phase are shown in Figs.~\ref{fig:HCB_chain}(a)-\ref{fig:HCB_chain}(c), obtained from tDMRG simulations
\cite{Vidal2004,White2004,Daley2004}.
We use a time step of $dt = 0.04/J$ and a bond dimension of up to $\chi = 1000$ and keep the discarded weight in each time step under $10^{-10}$. 
The disorder average is performed over about $500$ realizations.
These profiles show a crucial difference between the dynamics in the localized and the delocalized regime.
Deep in the localized regime, Fig.~\ref{fig:HCB_chain}(c), similar to the noninteracting models discussed
before, the density profiles quickly become stationary with an exponential decay even close to $j=0$.
In the ergodic phase, however, the density profiles never become stationary on the simulated time scales and for the 
values of interactions considered here. For $W=0.5J$ shown in Fig.~\ref{fig:HCB_chain}(a), the particles spread over the whole
considered system. 
Remarkably, we find a regime of slow dynamics \cite{BarLev2015,Vosk2015,Potter2015,Luitz2016,Luitz2016b} at intermediate disorder $W < W_c$ in Fig.~\ref{fig:HCB_chain}(b), where there
seems to persist an exponential decay of $n_j$ at finite times, but with a continuously growing $\xidw(t)$.
We note that  $\xidw(t)$ at the shortest time scales is on the order of the single-particle localization length. 
An explanation can thus be obtained in this  picture:
On short time scales, single particles can quickly expand into the
right, empty side within the single-particle localization length, thus leading to the exponential form of $n_j$. 
The interaction comes into play by scattering events at 
larger times, ultimately allowing the expansion over the whole system for infinite times. 

The slow regime is also reflected in the quantities $\VARn$ and $\Delta N$ in Figs.~\ref{fig:HCB_chain}(d) and \ref{fig:HCB_chain}(e), which behave
qualitatively in the same way. 
While both quantities saturate for $W > W_c$ and the results hardly differ from the noninteracting case shown by the dotted lines,
the slow growth becomes evident for $W \lesssim W_c$ at the intermediate time scales accessible to us.
The slow growth of both $\VARn$ and $\Delta N$ is, for $W \lesssim W_c$, the best described by (yet hard to distinguish from a power-law)
\begin{equation}
	\Delta N (t), \mbox{VAR}_n(t) = C(W) \log(tJ) + \mathrm{const}\,. \label{eq:fitDN}
\end{equation}
This growth is qualitatively different from the non-interacting case, where a saturation sets in after a faster initial increase.
The inset of Fig.~\ref{fig:HCB_chain}(e) shows the prefactor $C(W)$ extracted from a fit to the data of $\Delta N(t)$
for $t J > 10$. This allows us to extract $W_c$ since $C(W>W_c) = 0$ for the stationary profiles in the localized phase.
Our result for $W_c$ is compatible with the literature value $W_c/J = 3.5\pm 1$ \cite{Luitz2015,Pal2010,BarLev2015,Bera2015} (dashed line in the inset of Fig.~\ref{fig:HCB_chain}).

\begin{figure}
\includegraphics[width=\columnwidth]{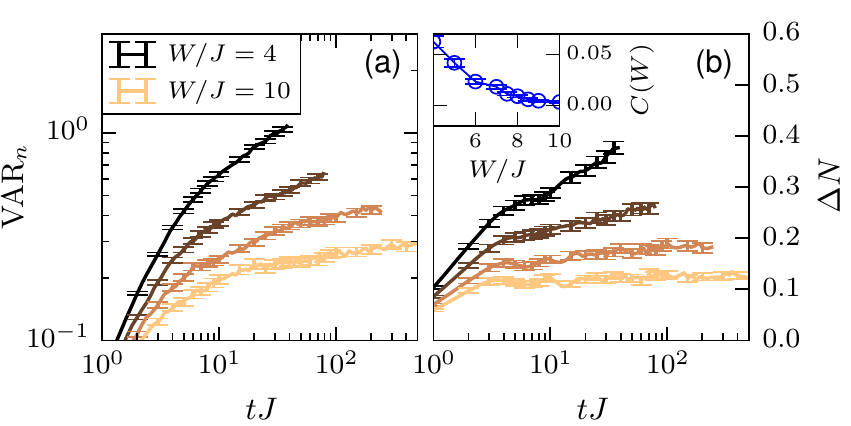}
\caption{tDMRG results for a two-leg ladder ($L=60$) of interacting spinless fermions with $V=J$. 
	(a) Variance $\VARn$  at $W/J=4, 6, 8, 10$ (top to bottom).
	(b) Number of emitted particles $\Delta N(t)$.
	Inset: $C(W)$ from fit to Eq.~\eqref{eq:fitDN} for $t J > 10$.
}
\label{fig:SLF_ladder}
\end{figure}

{\bf Interacting fermions on a ladder.}
As a first step towards 2D systems, we present results for the dynamics of
interacting spinless fermions on a two-leg ladder in the presence of diagonal disorder.
The simulations are done with a variant of tDMRG suitable for long-range interactions \cite{Zaletel2014},
with a time step $dt = 0.01/J$.
Figures \ref{fig:SLF_ladder}(a) and \ref{fig:SLF_ladder}(b) show the variance $\VARn$ and $\Delta N$ for $V=J$, respectively. 
As for the chain, we observe that both the variance and $\Delta N$ have a tendency to saturate for large disorder strength, while they keep growing for small disorder.
The data are best described by Eq.~\eqref{eq:fitDN} and we extract $C(W)$  from  fits of the data for $tJ > 10$ to Eq.~\eqref{eq:fitDN}. 
The results of these fits shown in 
the inset of Fig.~\ref{fig:SLF_ladder} suggest a critical disorder strength $8 \lesssim W_c/J \lesssim 10$, in good agreement with 
the value of  $W_c/J= 8.5 \pm 0.5$ found in an ED study of the isotropic Heisenberg model on a two-leg ladder \cite{Baygan15}
(the two models differ by correlated hopping terms which are not believed to be important for the locus of the transition).

{\bf Summary and outlook.}
We analyzed the domain-wall melting of fermions in the presence of diagonal disorder, motivated by
a recent experiment \cite{Choi16} that was first in using this setup for interacting bosons in 2D.
Our main result is that several quantities accessible to experimentalists (such as the number of propagating particles and the variance of their particle density) 
are sensitive to localization
and can be used to locate the disorder-driven metal-insulator transition, based on
our analysis of several models of noninteracting and interacting fermions for which the phase diagrams are known.
Notably, this encompasses a two-leg ladder as a first step towards numerically simulating the dynamics of 
interacting systems with disorder in the 1D-2D crossover.
Our work further indicates that care must be taken in extracting quantitative results from finite systems or
finite times since the approach to the stationary regime can be slow. Interestingly, we observe a slow dynamics
in the ergodic phase of interacting models as the transition to the MBL phase is approached,  which deserves further investigation.

The domain-wall melting thus is a viable approach for  theoretically and experimentally studying 
disordered interacting systems, and we hope that our work will influence future experiments on quasi-1D
systems where a direct comparison with theory is feasible. Concerning 2D systems, where numerical simulations of real-time dynamics face severe limitations, 
our results for two-leg ladders provide confidence that the domain-wall melting 
is still a reliable detector of localization as well, as evidenced in the experiment of \cite{Choi16}. 
Even for clean systems, experimental studies of domain-wall  melting in the presence of interactions could provide 
valuable insights into the nonequilibrium transport properties of interacting quantum gases \cite{Antal1999,Gobert2005,ronzheimer13,Vidmar2013,
schneider12,Hauschild2015}.
For instance, even for the isotropic spin-1/2 chain ($V=1$ in our case),
the qualitative nature of transport is still an open issue
\cite{Herbrych2011,Grossjohann2010,Znidaric2011,Karrasch2013,Sirker2011,Bertini2016, prosen13,steinigeweg14}.
Moreover, the measurement of diffusion constants would be desirable \cite{Karrasch2014}.

{\bf Acknowledgments.}
We thank I.~Bloch, J.~Choi, G.~De~Tomasi, and C.~Gross for useful and stimulating discussions.
F.P. and F.H.-M. were supported by the DFG (Deutsche Forschungsgemeinschaft) Research Unit FOR 1807 through Grants No.~PO 1370/2-1 and No.~HE 5242/3-2.
This research was supported in part by Perimeter Institute for Theoretical Physics. Research at Perimeter Institute is
supported by the Government of Canada through Industry Canada and by the Province of Ontario through the Ministry of
Economic Development \& Innovation.

\bibliography{references}


\setcounter{figure}{0}
\setcounter{equation}{0}
\renewcommand{\thefigure}{S\arabic{figure}}
\renewcommand{\theequation}{S\arabic{equation}}
\section*{\large Supplemental Material}

\input{suppmattext}

\end{document}

%% file: suppmattext.tex
\renewcommand{\thefigure}{S\arabic{figure}}
\renewcommand{\theequation}{S\arabic{equation}}

\begin{figure}[tb]
\includegraphics[width=\columnwidth]{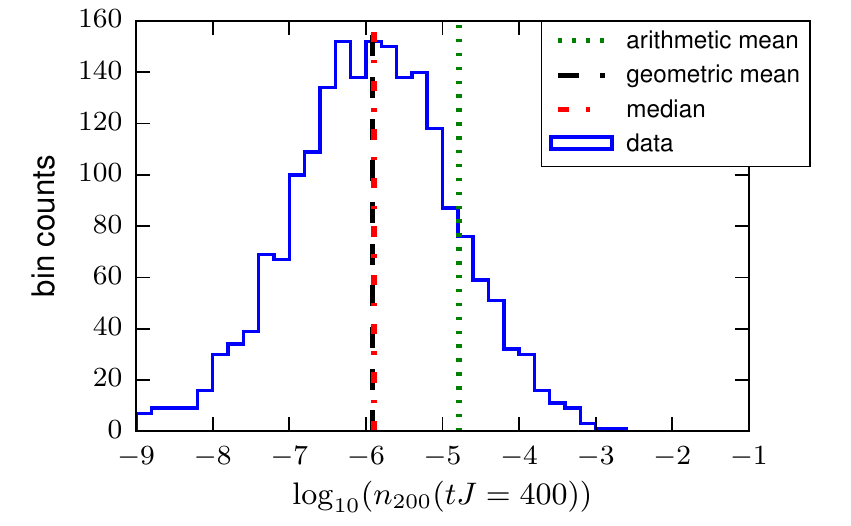}
\caption{Distribution of $n_j$ on a logarithmic scale, exemplary for the free fermions at site $j=200$, $tJ = 400$ and $W=0.5J$.
}
\label{fig:histogram}
\end{figure}
\subsection{Disorder Statistics}

An exemplary distribution of $n_j$ for the free fermion case is shown in Fig.~\ref{fig:histogram}.
In a rough approximation, the probability for a particle to hop the $j$ sites out of the domain wall 
can be seen as a product of the hopping probabilities to neighboring sites, which depend on the specific disorder
realization. 
The geometric mean $\bar{n}_j$ is thus a natural choice for the average over different disorder realizations.
As evident from Fig.~\ref{fig:histogram}, it coincides with the median and represents the typical value. In contrast, the
arithmetic mean is an order of magnitude larger as it puts a large weight in the upper tail of the distribution.

Although the geometric mean is a good choice for $n_j$, it is reasonable to use the arithmetic mean for other
quantities such as $\VARn$ and $\Delta N$: they represent quantities integrated over $j$ for a given disorder
realization. We checked that the arithmetic mean is close to typical values for these quantities.

\subsection{Aubry-Andr{\'e} model}

\begin{figure}[tbh]
\includegraphics[width=\columnwidth]{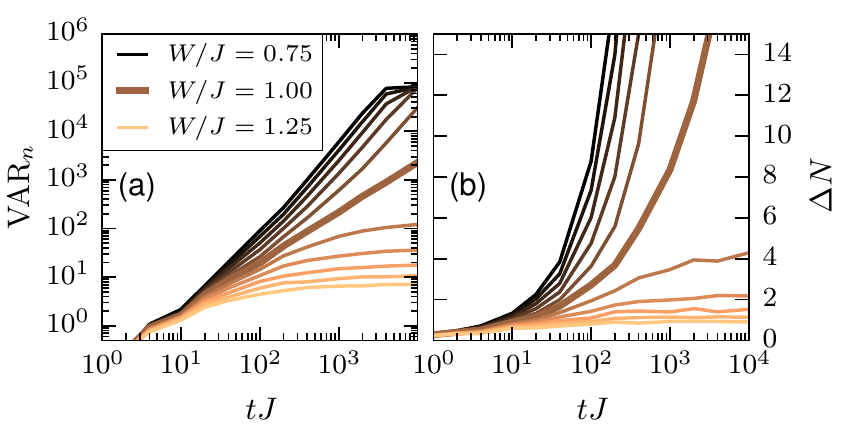}
\caption{ED results ($L =2000$) for the Aubry-Andr{\'e} model with a localization transition at
	$W_c=J$ (indicated by the thick lines). 
	(a) Variance VAR$_{n}$ for $W/J=0.75, 0.8, 0.85, \dots, 1.25$ (top to bottom). 
	(b) Number of emitted particles $\Delta N(t)$.
	Error bars are smaller than symbol sizes and omitted.
}
\label{fig:AA}
\end{figure}

We now focus on the dynamics in the Aubry-Andr\'{e} model, where a quasiperiodic
modulation is introduced in Eq.~\eqref{eq:ham} via $\epsilon_j = W \cos(2 \pi r j + \phi_0)$ (employed in 
the MBL experiments of \cite{Schreiber2015,Bordia2015}).
We set the irrational ratio $r$ to  $r= (\sqrt{5}-1)/2 = 0.61803\dots$ and
perform the equivalent to disorder averages by sampling over the value of the phase $\phi_0 \in [0,2\pi)$.
This noninteracting model has a delocalization-localization transition at $W_c/J=1$, where the single-particle localization length
diverges as $\xi^{(1)}_{\rm loc} = \log\left(\frac{W}{J}\right)$ \cite{Aubry80}.
Similar to the previously considered Anderson model, the density profiles become stationary with an exponential tail in the
localized phase for $W > W_c$.
As $W$ is varied, a clear transition is visible in the time dependence of both $\VARn$ and $\Delta N$ shown in Figs.~\ref{fig:AA}(a) and \ref{fig:AA}(b), respectively, which become stationary
for $W > W_c$, while growing with a power law for $W < W_c$.
The corresponding domain-wall decay length $\xidw$ (see the  inset of Fig.~\ref{fig:nonint}(b)) diverges 
as $W_c$ is approached from above, in excellent agreement with the single-particle localization
length of that model \cite{Aubry80}.
The maximum value of $\xidw$ in the extended phase reached at long times diverges with $L$.

